\documentclass[reprint,amsmath,amssymb,aps, prl, balancelastpage, nolongbibliography, floatfix,superscriptaddress,nofootinbib]{revtex4-2}

\usepackage{bm}
\usepackage{hyperref}
\usepackage{xcolor}
\usepackage{hyperref}
\hypersetup{
  colorlinks=true,
  linkcolor= [rgb]{0,0.18,0.65}, 
  linkbordercolor=black, 
  citecolor=[rgb]{0,0.18,0.65},
  urlcolor = [rgb]{0,0.18,0.65}
  }
\usepackage{graphicx,epsfig,chemarr}
\usepackage{times}

\usepackage{eucal}
\usepackage{amsmath,amssymb}
\usepackage{relsize}
\usepackage{physics}
\usepackage{comment}
\usepackage{siunitx}
\DeclareSIUnit\molar{\mole\per\cubic\deci\metre}
\DeclareSIUnit\Molar{\textsc{m}}

 \def\r{\right}
 \def\l{\left}
 \def\lang{\left\langle}
 \def\rang{\right\rangle}
\def\Rb{\mathbf{R}}
\def\rb{\mathbf{r}}
\def\vb{\mathbf{v}}

\def\mb{\mathbf{m}}
\def\sS{{\mathsmaller{\mathrm{S}}}}
\def\sD{{\mathsmaller{\mathrm{D}}}}
\def\sS{{\mathsmaller{\mathrm{D}}}}
\def\sT{\mathrm{tot}}

\def\vfd{\vb_\sD}
\def\Fb{\mathbf{F}}
\def\fb{\mathbf{f}}
\def\lb{\boldsymbol{\ell}}
\def\lbs{\lb_\sS}

\def\lsS{\ell_\sS}

\def\Rtr{a} 
\def\Str{{\mathrm{str}}}
\def\Rot{{\mathrm{rot}}}
\def\hrb{\hat{\rb}}
\def\hrbs{\hat{\rb}}

\def\heb{\hat{\mathbf{e}}}
\def\hfb{\hat{\mathbf{f}}}
\def\hub{\hat{\mathbf{u}}}
\def\zero{{\mathsmaller{\mathbf{0}}}}
\def\sal{{\mathsmaller{\mathbf{\alpha}}}}
\def\sbe{{\mathsmaller{\mathbf{\beta}}}}

\def\hubz{\hub_\zero}
\def\hubzz{\hub_{\zero z}}
\def\huba{\hub_\sal}
\def\hubb{\hub_\sbe}
\def\Omegab{\mathbf{\Omega}}

\def\d{\mathrm{d}}
 \def\tauc{\tau_{\mathrm{c}}}
 \def\taur{\tau_{\mathrm{r}}}
 \def\taum{\tau_{\mathrm{m}}}
 \def\taud{\tau_{\mathrm{s}}}
 
 \def\tauef{\tau_{\mathsmaller{\mathrm{eff}}}}
 
 \def\taup{\tau_{\mathrm{p}}}
 \def\kcat{k_{\mathrm{cat}}}
 \def\Drs{D_{\mathrm{r}}}
 \def\Dth{D_{\mathrm{th}}}
 \def\Dfd{D_{\mathrm{fd}}}
 \def\nth{\eta_{\mathrm{th}}}

 \def\Dtr{\mathbf{D}^{\mathrm{tr}}}
 \def\mpara{m_{\mathsmaller{\parallel}}}
 \def\mparai{m_{\mathsmaller{\parallel}i}}
 \def\mperp{m_{\mathsmaller{\perp}}}
 \def\mperpi{m_{\mathsmaller{\perp}i}}
 \def\mtot2{m_{\mathrm{s}}^2}
 
 \def\mtots{m_{\mathrm{s}}}
 \def\fpara{f_{\mathsmaller{\parallel}}}
 \def\fperp{f_{\mathsmaller{\perp}}}
 \def\Fpara{\Fb_{\mathsmaller{\parallel}}}
 \def\Fperp{\Fb_{\mathsmaller{\perp}}}
\def\Fparaa{\Fb_{{\mathsmaller{\parallel}}\alpha}}
\def\Fparab{\Fb_{{\mathsmaller{\parallel}}\beta}}
\def\Fperpa{\Fb_{\mathsmaller{\perp}\alpha}}
\def\Fperpb{\Fb_{\mathsmaller{\perp}\beta}}
 \def\sB{{\mathsmaller{\mathrm{B}}}}
 \def\kB{k_\sB}
 
 \def\kBT{\kB T}

\def\eg{{\it e.g.},~}

 \def\SF{S_{\mathsmaller{\mathrm{F}}}}
 \def\ST{S_{\mathsmaller{\mathrm{T}}}}
 
 \def\Pb{\mathbf{P}}
 \def\lmc{\lambda_1}
 \def\lmp{\lambda_2}

 \def\Csub{c_{\mathsmaller{\mathrm{s}}}}
 \def\Km{K_{\mathsmaller{\mathrm{m}}}}

\def\GG{\mathbf{\mathcal{G}}}
\def\EE{\mathbf{\mathcal{E}}}
\def\Gab{\GG_{\alpha\beta}}
\def\sixth{{\textstyle\frac{1}{6}}}

\def\third{{\textstyle\frac{1}{3}}}

\definecolor{YKB}{rgb}{0.00,0.18,0.65}

\begin{document}

\title{Gauging nanoswimmer dynamics via the motion of large bodies}

\author{Ashwani Kr. Tripathi}
 \affiliation{Center for Soft and Living Matter, Institute for Basic Science (IBS), Ulsan, 44919, Republic of Korea}
\author{Tsvi Tlusty}%
\email{tsvitlusty@gmail.com}
\affiliation{Center for Soft and Living Matter, Institute for Basic Science (IBS), Ulsan, 44919, Republic of Korea}%
\affiliation{Department of Physics, Ulsan National Institute of Science and Technology, Ulsan, 44919, Republic of Korea}
\affiliation{Department of Chemistry, Ulsan National Institute of Science and Technology, Ulsan, 44919, Republic of Korea}

\date{\today}

\begin{abstract}
Nanoswimmers are ubiquitous in bio- and nano-technology but are extremely challenging to measure due to their minute size and driving forces. A simple method is proposed for detecting the elusive physical features of nanoswimmers by observing how they affect the motion of much larger, easily traceable particles. Modeling the swimmers as hydrodynamic force dipoles, we find direct, easy-to-calibrate relations between the observable power spectrum and diffusivity of the tracers and the dynamic characteristics of the swimmers---their force dipole moment and correlation times. 
 
\end{abstract}

\maketitle

\noindent\emph{Introduction.---}
In recent years, nanoscale swimmers attracted much interest as a basic physical phenomenon with promising potential in biomedical and technological applications~\cite{ricotti2017, Sanchez2015, Sanchez2021}. Examples include artificial swimmers, such as chemically powered nanomotors~\cite{Gao2014,Sanchez2015, Zhang2021,Wang2013, Paxton2004,Geoffrey2005,Geoffrey2010,Ghosh2018}, bio-molecules that exhibit enhanced diffusion ~\cite{Jee2018}, and bio-hybrid swimmers~\cite{ricotti2017, Sanchez2021, Ma2016, Patino2018}. 
Because of their minute size, the motion of nano- and micro-swimmers is deep in the low-Reynolds regime where viscous forces dominate over inertia~\cite{Bechinger2016,Gompper2015,HStrak2016}. But the swimmers also experience stochastic forces from the surrounding solvent molecules, and at the nanoscale, these thermal fluctuations become comparable to the typical driving forces. Hence, measuring the properties of nanoswimmers using traditional techniques, such as fluorescence correlation spectroscopy~(FCS) and dynamic light scattering~(DLS)~\cite{Jee2018,Jee2018a,Zhang2021}, is extremely difficult, leaving core questions in the field---particularly, whether enzymes and small catalysts self-propel---open and a matter of lively debate~\cite{Wang2020,Tian2021,Xu2019,Xu2020,Chen2020, Agudo2018,Ghaleh2022,Lucy2021}. 

An alternative path to characterize nanoswimmers is by observing how they affect the motion of large, micron-size particles~\cite{Zhao2017,Lee2014}, such as silica beads~\cite{Guo2014}, or vesicles~\cite{Ahmed2018}. 
 Tracer motion has been extensively investigated in suspensions of \emph{micro}-swimmers, especially microbes~\cite{Libchaber2000,Goldstein2009,Mino2011,Poon2013, Morozov2022,Cheng2016,cheng2016A,Eremin2021,soto2013,Marenduzzo2014,Kasyap2014,childress2011,Ortileb2019,Kanazawa2020} whose size is comparable to the spherical and ellipsoidal tracers used. These studies typically report a many-fold enhancement of the tracers' diffusion compared to thermal diffusion~\cite{Libchaber2000,Goldstein2009,Mino2011,Poon2013, Morozov2022,Cheng2016,cheng2016A,Eremin2021}. Theoretical models that explain the observed enhancement are based on the hydrodynamic interactions induced by the swimmers' motion over long, persistent trajectories~\cite{soto2013,Marenduzzo2014,Kasyap2014,childress2011}. The enhancement is proportional to the volume fraction of swimmers, their self-propulsion speed~\cite{Goldstein2009,Mino2011,Poon2013}, and geometrical factors, such as the average run-length of the swimmer before it changes direction~\cite{soto2013,Marenduzzo2014,Kasyap2014,childress2011}. Here, thermal fluctuations have a negligible effect compared to self-propulsion, as indicated by a many-fold increase in diffusivity.       
 
 While similar experimental studies of tracer motion in a suspension of \emph{nano}-swimmers are much fewer~\cite{Zhao2017,Lee2014}, they suggest a common mechanism of momentum transfer from swimmers to tracers which may operate at the molecular scale of organic reactions~\cite{Wang2020,Tian2021, Dey2016}. Unlike the motion of the nano-swimmers, the motion of the micro-sized tracer particles is easy to track, for example, by video microscopy, and one could therefore, in principle, use tracers to probe the forces generated by the swimmers. But the application of this potentially advantageous method is hindered by the lack of understanding and rigorous computation of the hydrodynamic coupling between nano-swimmers and tracers.

Here, we present a first-principles theory that addresses this problem by linking the tracer’s motion to the dynamics of the swimmer suspension. The theory derives the hydrodynamic flow field generated by swimmers, and its effect on the tracer's motion, accounting for three physical effects dominant in the nano-regime: (a) Thermal fluctuations -- due to their nanometric size, the swimmers are subjected to strong thermal forces, giving rise to vigorous stochastic rotation and translation. (b) Stochastic driving -- nano-swimmers are often propelled by strongly fluctuating chemical reactions, where intermittent activity bursts are separated by rest periods as, for example, in enzymatic reactions. (c) Near-field hydrodynamics: a micron-sized tracer is effectively a large-scale boundary and swimmers are in the near-field view of the tracer.  
By computing these three physical effects~(see Model section), we obtain our main results: simple expressions for the observable force-force autocorrelation, power spectrum, and diffusivity of the tracer particles from which one can gauge the nano-swimmer's dipole moment and persistence time (particularly, Eqs~(\ref{eq:RelDtr},\ref{eq:SF},\ref{eq:RelDtr1})).

In the following, we explain the underlying physical intuition and main steps of the derivation (whereas the details are given in the Supplemental Material~(SM)~\cite{SM}). We then perform a Brownian dynamics simulation of the tracer motion in the swimmer suspension and compare it with our analytical findings. Finally, we propose and demonstrate, using the simulation results, how to use the derived estimates in experiments, especially as physical bounds for testing hypothesized self-propulsion mechanisms.\\

\noindent\emph{Model.---} The swimmer motion is within the highly viscous regime, at low Reynolds number, and it is force- and torque-free. Hence, the leading order contribution to the flow field of a swimmer is due to the force dipole~\cite{Lauga2009,Ishikawa2009, kim2005,pozrikidis1992}.
Thus, we consider the nano-swimmer suspension as an ensemble of force dipoles, each consisting of two equal and opposite point forces (Stokeslets) of strength $\fb = f\hfb$, separated by an infinitesimal distance $\lbs = \lsS\heb$. The resulting force dipole, which determines the far-field flow is the tensor $\mb_{\alpha\beta} = m \heb_{\alpha}\hfb_{\beta}$, where $m = f \lsS$ is the dipole's magnitude. The velocity field induced at a distance $\rb$ from the dipole is obtained from the gradient of  $\Gab$, the hydrodynamic Green function, $\vfd^{\alpha}(\rb) = \mb_{\beta\gamma}\partial_{\gamma}\Gab(\rb)$. The Green function, $\Gab = (\delta_{\alpha\beta}+\hrb_{\alpha}\hrb_{\beta})/(8\pi\eta r)$, is the mobility tensor, defined as the flow generated by a Stokeslet of unit strength. 
From a physical point of view, it is instructive to divide the induced velocity $\vfd$ into symmetric and anti-symmetric parts, 
\begin{align}
\label{eq:VelFD}
  \vfd(\rb) = \frac{\l[ 3\l(\fb\vdot\hrb \r) \l(\lbs\vdot\hrb \r)-\fb\vdot\lbs \r] \hrb}{8\pi\eta r^2}  +
  \frac{\l(\fb\cross\lbs \r)\cross\hrb}{8\pi\eta r^2}~.
\end{align}
 The first, symmetric contribution, called \textit{stresslet}, arises from the straining motion of the dipole when the forces are parallel to the dipole's orientation. The second, anti-symmetric term, known as \textit{rotlet}, corresponds to the rotational motion of the dipole, arising when the forces and the dipole are not aligned~\cite{batchelor1970,blake1974,chwang1974, chwang1975}.

A tracer subjected to the velocity field $\vfd$ experiences a hydrodynamic drag that depends on its size. Tracers are much larger than the swimmers, move much slower, and therefore effectively serve as static boundaries. Finding the flow field near the tracer's surface generally requires calculating the image system of the force dipole by a multipole expansion~\cite{kim2005,blake1971,pozrikidis1992}. However, one finds that the force $\Fb$ exerted on a spherical tracer depends only on the leading-order monopole term, and can be calculated using Fax\'en's law~\cite{happel1965, rallison1978},
\begin{equation}
 \label{eq:FaxenM}
     \Fb = \eval{ 6\pi\mu \Rtr\l(1+\sixth \Rtr^2\laplacian \r)\vfd(\rb) }_{\rb=0}~,
 \end{equation}
where $\Rtr$ is the radius of the spherical tracer whose center is at $\rb=0$. We notice that the second term arises from the large scale of the tracer and becomes negligible when the tracer size is small compared to its distance from the swimmer~($\Rtr \ll r$). Substituting the dipolar velocity field (Eq.~(\ref{eq:VelFD})) in Fax\'en's law (Eq.~(\ref{eq:FaxenM})), we obtain the force exerted on the tracer, $\Fb = \Fb_{\Str} + \Fb_{\Rot}$,
where the contributions from the stresslet and rotlet are,
\begin{align}
\label{eq:FcolSSM}
&\Fb_{\Str} = \frac{3\Rtr m}{4r^2}
\Bigg[\hrb\l(-\heb\vdot\hfb +3(\heb\vdot\hrb)(\hfb\vdot\hrb)\r)+ 
\nonumber \\
& \frac{\Rtr^2}{r^2}\Big[\hrb\l(\heb\vdot\hfb  -5(\heb\vdot\hrb)(\hfb\vdot\hrb)\r) + \heb(\hfb\vdot\hrb)+ \hfb(\heb\vdot\hrb)\Big]\Bigg]~,
\nonumber\\
    &\Fb_{\Rot} = \frac{3\Rtr m}{4r^2}\left[\hfb(\heb\vdot\hrb)-\heb(\hfb\vdot\hrb)\right]~,
\end{align}
and $r$ is the distance between the swimmer and sphere center with the unit vector $\hrb$.

Consider a suspension consisting of an ensemble of $N$ force dipoles of strengths $\{m_i\}$ located at positions $\{\rb_i\}$ with dipole and force orientations, $\{\heb_i\}$ and $\{\hfb_i\}$. Due to the linearity of the Stokes flow and the minute size of the swimmers, one can neglect higher-order terms and multiple-scattering interactions among the dipoles. Within this approximation, the total force on the tracer is simply a superposition of the forces exerted by the individual dipoles, 
\begin{equation}
\label{eq:Ftotal}
    \Fb^{\sT} = \sum_{i=1}^N\l(\Fb^{(i)}_{\Str} + \Fb^{(i)}_{\Rot}\r)~,
\end{equation}
where $\Fb^{(i)}_{\Str}$ and $\Fb^{(i)}_{\Rot}$ are the contributions due to stresslet and rotlet from the $i$-th dipole.

Due to their nanometric size, the dipoles experience strong stochastic kicks by the solvent molecules and other noise sources present in the suspension, resulting in two effects. First, the fluctuations induce diffusive motion, translational and rotational. 
The translational diffusivity scales inversely with the swimmer's size~, $D = \kBT/(3\pi\eta \lsS)$, whereas rotational diffusion has an inverse cubic dependence,~$\Drs = \kBT/(\pi\eta \lsS^3)$. As a result, during a typical rotational timescale $\taur = 1/(2\Drs)$, a particle will diffuse to a distance ${\sim}\lsS$ while rotating about one radian. Since the separation between tracers and dipoles is typically much larger than the dipole size~($r\gg \lsS$), the change in dipole position due to translational diffusion has a negligible effect on the hydrodynamic force it exerts on the tracer~(Eq.~\ref{eq:FcolSSM}). In contrast, within the same period, a swimmer performs, on average, a full rotation, thus strongly affecting the force on the tracer. This stochastic wandering of the orientations $\heb$ and $\hfb$ on the surface of a unit sphere is captured by a rotational diffusion equation~\cite{doi1988,doi2013,berne2013}, from which we obtain probability moments for orientations that are required for the calculating moments of the force $\Fb^{\sT}$ (see details in SM~\cite{SM}).

The second stochastic effect stems from internal fluctuations of the force dipole that vary its magnitude, $m(t)$~\cite{mikhailov2015}. Certain force dipoles, particularly those fueled by chemical cycles, will work in bursts with finite persistence time $\tauc$ each stroke. Such swimmers are additionally characterized by $\tau_p$, the typical cycle period between strokes, which in catalysts is the inverse of the turnover rate. We take for simplicity, the bursts of duration $\tauc$ through which the force dipole is constant, $m(t) = m$. These square bursts occur on average every $\taup$, and in between the bursts, the force dipole is idle $m(t) = 0$. The resulting autocorrelation of the moment is 
\begin{equation}
    \lang m(t)m(0) \rang =
   m^2 b  \l[b + \l(1- b \r) e^{-t/\taum} \r]~,
\end{equation}
where $b=\tauc/\taup$ is the relative fraction of the bursts during the cycle, and $\taum=\tauc (1-b)$ is the timescale of moment fluctuations~(See SM~\cite{SM}).\\

\noindent\emph{Results.---}~ The fluctuations in the dipoles' orientation and moment render the force $\Fb^{\sT}$ stochastic and to calculate its statistics, we consider an arbitrarily oriented force dipole with axisymmetric~($\heb=\hfb$) and transverse~($\heb\perp\hfb$) components, $\fb =\fpara\heb+\fperp \hfb$. For the axisymmetric dipole, only the stresslet contributes to the total force, whereas for the transverse dipole both stresslet and rotlet contribute~(Eqs.~(\ref{eq:FcolSSM}, \ref{eq:Ftotal})).  Thus the total force can be expressed as the sum of 
 axisymmetric and transverse components, $\Fb^{\sT}(t) = \Fpara^{\sT}(t) + \Fperp^{\sT}(t)$).
To find the mean force and its fluctuations, we average over dipole positions, orientations, and moments, as detailed in the SM~\cite{SM}. We find that the net mean force vanishes, $\lang\Fb^{\sT} (t)\rang = 0$, as expected from symmetry. Thus, the surviving dominant moment is the force auto-correlation,
\begin{equation}
\label{eq:FFCorr1}
\lang\Fb_\alpha^{\sT}(t) \Fb_\beta^{\sT}(0)\rang =  \delta_{\alpha\beta}\frac{2\pi}{5} c_0 \Rtr \mtot2 b\l[ be^{-\frac{3t}{\taur}}+(1-b)e^{-\frac{t}{\taud}}\r]~,
\end{equation}
where the Greek indices denote hereafter Cartesian components and $c_0=\lang\sum_i\delta(\rb-\rb_i)\rang$ is average concentration of dipoles. $\mtot2 \equiv \mpara^2 + 2 \mperp^2$, is the effective squared momentum of the swimmer with the axisymmetric and transverse contributions, $\mpara=\fpara \lsS$ and $\mperp = \fperp \lsS$, and the effective timescale of the swimmer $\taud$ is the harmonic mean of the timescales of rotational motion and dipole-moment fluctuations, $\taud^{-1} = \taum^{-1} + 3\taur^{-1}$.

A chief measurable quantity is the diffusion coefficient for the tracer, which is computed using the Green-Kubo relation~\cite{Kubo1966,Zwanzig2001},
\begin{equation}
\label{eq:Diff}
    \Dtr_{\alpha\beta} = \frac{1}{\gamma^2}\int_0^{\infty}\lang\Fb_\alpha^{\sT}(t) \Fb_\beta^{\sT}(0)\rang \dd{t}~,
\end{equation}
where $\gamma = 6\pi\eta\Rtr$ is the friction coefficient of the spherical tracer, and $\eta$ is the viscosity of the suspension. Substituting the autocorrelation from Eq.~(\ref{eq:FFCorr1}) into the Green-Kubo relation, we obtain the isotropic diffusivity,
\begin{equation}
\label{eq:Dtr1}
\Dtr_{\alpha\beta}= \frac{\delta_{\alpha\beta}}{90\pi} \,\frac{c_0  \mtot2\tauef}{\eta^2\Rtr}~,
\end{equation}
where the timescale $\tauef= \third b^2\taur  + b(1-b)\taud$ is the linear combination of the timescales $\taur$ and $\taud$.
We see that similar to the thermal diffusivity of the tracer $\Dth = \kBT/\gamma = \kBT/(6\pi\eta\Rtr)$, suspension-induced diffusion is also inversely proportional to its size, allowing us to define size-independent relative enhancement,
\begin{equation}
\label{eq:RelDtr}
    \EE = \frac{\Dtr_{\alpha\alpha}}{\Dth}=\frac{c_0\mtot2 \tauef}{15\eta\kBT}~.
\end{equation}

Another experimentally measurable quantity is the power spectrum of force for tracer, $\SF(\omega)$, the Fourier transform of the force autocorrelation, which measures the frequency-dependent response of the hydrodynamic force $\Fb_\alpha^{\sT}(t)$,
\begin{equation}
    \SF(\omega) = \int_{-\infty}^{\infty} e^{i\omega t}\lang\Fb_\alpha^{\sT}(t) \Fb_\alpha^{\sT}(0)\rang \dd{t}~.
\end{equation}
From Eq.~(\ref{eq:FFCorr1}), we find that the power spectrum is a sum of two Lorentzians,  
\begin{equation}
\label{eq:SF}
    \SF(\omega) = \frac{4\pi}{5}c_0\Rtr\mtot2 b\l[b \frac{\third\taur}{\omega^2(\third\taur)^2+1} + (1-b)\frac{\taud}{\omega^2\taud^2+1}\r]~.
\end{equation}
The Lorentzians merge in the limit of fast rotation, $\taur \ll \taum$, typical to nanoswimmers, or when the swimmer self-propels continuously ($b=1$). The Fourier transform of the  tracer displacement, $\Rb(\omega) =  \int_{-\infty}^{\infty} e^{i\omega t}~\Rb(t) \dd{t}$, is connected to the power spectrum $\SF(\omega)$ via a fluctuation-dissipation relation,
\begin{equation}
\label{eq:FDTSF}
 \lang |\Rb(\omega)|^2\rang = |\chi(\omega)|^2\l[\SF(\omega)+\ST(\omega)\r]~,
\end{equation}
where $\ST(\omega) = 2\gamma\kBT$ is the power spectrum of thermal fluctuations and $\chi(\omega)$ is the response function, which is $\chi(\omega) = (i\gamma\omega)^{-1} = (6 i\pi\eta \Rtr\omega)^{-1}$ for a spherical tracer in a Newtonian fluid.\\

\begin{figure*}[!htbp]
\includegraphics[width=\textwidth]{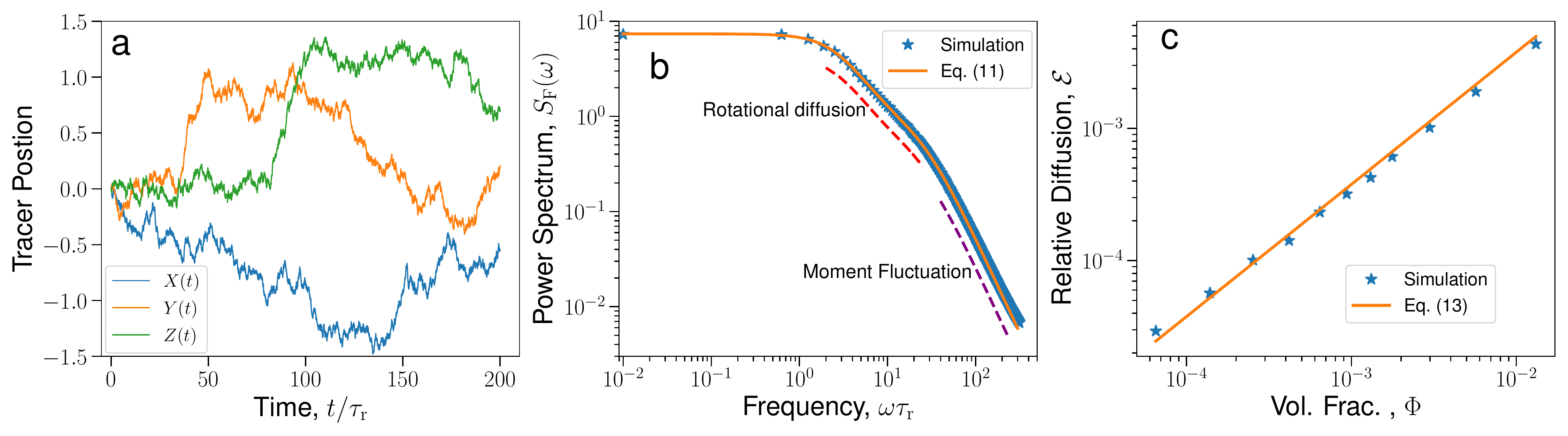}
\caption{\textbf{Numerical realization of tracer dynamics in a nano-swimmer suspension}. 
(a) A tracer trajectory $\Rb(t)=(X(t),Y(t),Z(t))$ in a typical simulation. 
(b) Force power spectrum $\SF(\omega)$ computed by averaging over tracer trajectories (see text). Blue stars are the numerical data, and the yellow line is the fit to the analytic expression in Eq.~(\ref{eq:SF}), for volume fraction of swimmers $\Phi=0.013$. 
(c) The variation of the relative enhancement of the diffusion $\EE$ (blue stars) with the volume fraction of the swimmers $\Phi$. 
Using the analytic model (Eqs.~\ref{eq:SF},\ref{eq:FDTSF}) we extracted from the trajectories the values of the nano-swimmer parameters,
which agree with the simulation parameters (see text).
}
\label{Fig1}
\end{figure*}

\noindent\emph{Numerical Simulation.---}
 To verify our analytical findings, and demonstrate how they can be used in experiments, we consider a numerical realization where dipoles are  randomly distributed in a cubic box with a tracer particle initially located at the box's center. The dipoles undergo rotational and translational Brownian motion, and their dynamics are modeled using the standard Langevin's framework. For clarity, we chose physical units where energy, length, and time are measured in the unit of thermal energy~($\kBT$), dipole length~($\lsS$), and the rotational timescale~($\taur$), respectively. The hydrodynamic interactions between dipoles and the tracer are calculated using Eq.~(\ref{eq:FaxenM}). Fluctuations in the dipole moments due to reaction stochasticity are modeled as a the two-state Markov process with the parameters $\mtots=10$, $\tauc=0.1$ and $\taup=0.2$. Thus, the relative fraction of bursts is $b=0.5$, and the effective time scale of moment fluctuations is $\taud=0.0435$ (See SM for more details).
 
During the simulation, we recorded the tracer's trajectory $\Rb(t)$, as one would typically measure in an experiment (Fig.~\ref{Fig1}a). We also recorded, independently, the total force exerted on the tracer by the nano-swimmer suspension as a function of time, $\Fb^{\sT}(t)$. By averaging over an ensemble of trajectories, we can calculate the diffusion coefficient of the tracer, $\Dtr = \lang|\Rb(t)|^2\rang /(6 t)$, and the force power spectrum $\SF(\omega)$ (from Eq.~\ref{eq:FDTSF}), which we verify against the power spectrum computed directly from $\Fb_{\alpha}^{\sT}(t)$ (Fig.~\ref{Fig1}b). In the graph of $\SF(\omega)$, one can easily notice two timescales of decay corresponding to the rotational diffusion and moment fluctuations. Taking into account finite-size effects (see details in SM), the power spectrum of the force, and diffusion coefficient of the tracer show excellent agreement with the numerical results (Fig.~\ref{Fig1}).
 
 We extract the properties of an individual dipole by fitting the numerical data for force power spectrum with Eqs.~(\ref{eq:SF}), corrected for the finite size effect. This provides us the values of average dipole moment $\mtots b= \num{5.56\pm 0.005}$, and of timescales $\taur=\num{0.98\pm 0.001}$ and $\taud=0.042 \pm 0.0002$. Thus, the time scale of momentum fluctuation $\taum$ can be estimated using $\taud^{-1} = \taum^{-1} + 3\taur^{-1}$, which yields $\taum=0.049$. Likewise, fitting of the diffusion coefficient data using Eq.~(\ref{eq:RelDtr1}) provides an estimate of the time scale ratio, $\tauef/\taur=\third b^2 + b(1-b)(\taud/\taur)$. Now, using the values of $\taur$ and $\taud$ extracted from the power spectrum fit, we obtain the relative fraction of bursts, $b\approx0.53$, the Markov process timescales, $\tauc = \taum/(1-b) \approx 0.1$, $\taup =\tauc/b \approx 0.2$, and the dipole moment, $\mtots \approx 10.5$ --
 all in agreement with the simulation parameters. This demonstrates how, in principle, the measurement of tracer trajectories allows the complete determination of the properties of an individual nanoswimmer. The above example represents the regime of $\taum \ll \taur$ or $\taum\approx \taur$. We also simulated the complementary regime, $\taum \gg \taur$, where the decay mode due to the moment fluctuations is masked by the rotational mode ($\taud\approx \taur$), but one can still extract the swimmer properties, particularly $\mtots$, $\taur$, and $b=\tauc/\taup$.\\

\noindent\emph{Physical interpretation and potential experimental use.---} The long-range hydrodynamic interactions link the motility of the tracer to the dynamics of an individual swimmer through simple testable relations (particularly, Eqs.~(\ref{eq:RelDtr}, \ref{eq:SF})). Thus, it bypasses the challenges of directly probing the nanometric objects and instead relies on the much easier measurement of larger objects using standard techniques. 
Within the superposition approximation (Eq. (\ref{eq:Ftotal}), the relative enhancement is \emph{independent} of properties of the passive tracers, especially their size $\Rtr$ (as long as they are spherical), but only reflects the specifications of the active swimmers. 

To gain further insight, consider a solution of roughly spherical swimmers of diameter $\lsS$. Substituting the volume fraction of the swimmers, $\Phi = (\pi/6) c_0 \lsS^3$, and their rotational diffusion time, $\taur = \pi\eta \lsS^3/(2\kBT)$, into Eq.~(\ref{eq:RelDtr}), we obtain an intuitive expression for the enhancement, 
\begin{equation}
\label{eq:RelDtr1}
        \EE = \frac{1}{5}\Phi \cdot \frac{\mtot2}{(\kBT)^2}\cdot \frac{\tauef}{\taur}~.
\end{equation}
Thus, the enhancement is proportional to the product of the volume fraction of swimmers and their squared momentum measured in $\kBT$ units. The timescale ratio, $\tauef/\taur=\third b^2 + b(1-b)(\taud/\taur)$, has two asymptotic limits: In the regime of fast rotation, $\taum\gg\taur$, corresponding to small swimmers, the ratio is  $\tauef/\taur=\third b$. In the other extreme of large, slowly-rotating swimmers, $\taum\ll\taur$, the ratio becomes $\tauef/\taur=\third b^2$. In the case of continuous propulsion, $b=1$, these asymptotic limits coalesce. For non-spherical swimmers, such as elongated swimmers in the shape of a prolate spheroid, Eq. (\ref{eq:Dtr1}) needs to be augmented by a geometric factor. 

The present results propose a straightforward recipe for probing biological or artificial nanoswimmers by observing the motility of large tracers using standard experimental techniques, such as confocal microscopy or optical tweezers~\cite{Huang2011,Govind2021}. From the observed positional data $\Rb(t)$ one can simply extract the mean square displacement $\lang|\Rb(t)|^2\rang$ and the resulting enhancement of the diffusivity, $\EE_{\rm exp} = \Delta D/\Dth = \lang|\Rb(t)|^2\rang /(6 \Dth t) - 1$. Then, using either of Eqs. (\ref{eq:RelDtr},\ref{eq:RelDtr1}), the measured $\EE_{\rm exp}$ is directly related to the product of the squared dipole moment of the swimmers (in $\kBT$ units) and their volume fraction.
This relative enhancement can also be detected as the relative widening of the probability distribution of a laser-trapped particle, expressed as the ratio of the variances, $\EE_{\rm exp} = \sigma^2/\sigma_{\rm th}^2 - 1$. Since the enhancement $\EE$ does not depend on the size of the spherical tracer (\eg Eq.~(\ref{eq:RelDtr1})), it can be calibrated using a series of tracers of various diameters. 

Another route to probing  the nanoswimmer dynamics is through the power spectrum $\SF(\omega)$. Using the fluctuation-dissipation relation (Eq.~(\ref{eq:FDTSF})), one can extract the power spectrum of the force fluctuations from the spectrum of tracer position, $\SF^{\rm exp}(\omega) = \lang |\Rb(\omega)|^2\rang/|\chi(\omega)|^2 - \ST(\omega)$. This experimental spectrum can then be compared to the theoretical double Lorentzian form (Eq.~(\ref{eq:SF})). In principle, this would allow to extract the dipole moment and the timescales associated with rotational and moment fluctuation of the swimmers, 
as demonstrated above in the numerical simulation. 
The same procedure can be applied to trapped particles, with a modified response function $\chi(\omega) = (i\gamma\omega + \kappa)^{-1}$, where $\kappa$ is the effective spring constant of the trap. 

As an example of potential use in experiments, we examine the diffusivity of tracers immersed in an enzymatic suspension. Within the hypothesis that enzymes are active nanoswimmers~\cite{Jee2018}, we assume that they exert dipole moment bursts of average duration $\tauc$ during the catalytic cycle, which are intermitted by idle periods of average duration $\taup-\tauc$, where $\taup$ is the average period of the enzymatic cycle. The Michaelis-Menten catalysis rate is $1/\taup = \kcat\Csub/(\Km+\Csub)$, where $\Csub$ is the substrate concentration, $\Km$ is the Michaelis constant, and $\kcat$ is the maximal turnover rate. Thus, we find that the relative fraction of bursting time is $b=\tauc/\taup=(\tauc \kcat)\cdot\Csub/(\Km+\Csub)$, reaching a maximum $b_{\rm max}=\tauc\kcat$ at saturation. Due to their size, the enzymatic nanoswimmers are in the fast rotation regime, $\taur \ll \taum$, where the relative enhancement is linear in $b$, and thereby follows the Michaelis-Menten saturation curve as a function of substrate concentration $\Csub$, as observed in experiments~\cite{Zhao2017}. 

Considering a suspension of enzymes whose dipole moment is $\mtots \approx \SI{10}{\kB}T$, where the typical distance between neighboring enzymes is about tenfold their size, $\Phi\approx \num{e-3}$, we find modest enhancement of tracer diffusion, $\EE\approx \SI{1}{\percent}$ (for $b=1$). However, the enhancement increases as the moment squared, $\EE \sim \Phi \mtot2$, so one expects a much more significant effect for larger, stronger swimmers (at the same volume fraction $\Phi$), for example, in artificial nanoswimmers, such as Au-Pt Janus particles~\cite{Lee2014}. At the other extreme, of molecular catalysts~\cite{Wang2020,Tian2021}, we still expect measurable effects on tracers in dense reactant solutions. On the other hand, observation of enhanced diffusion at much lower swimmer concentrations may indicate stronger force dipoles. One tentative speculation is that such strong forces arise when momentum and energy are channeled from electronic degrees of freedom of the reactants, which are fast and localized, to slow, collective modes of surrounding solvent molecules and ions~\cite{Jee2018}.
Similar collective effects are expected to show in the diffusivity of the nanoswimmers themselves. For example, the diffusivity of Janus particles should slightly increase with their concentration, beyond the self-enhanced diffusion of an isolated nanoswimmer. This hypothesis may also be tested in enzyme solutions, where we speculate that such long-range hydrodynamic interactions may also accelerate enzymatic kinetics by affecting the crossing rate of energy barriers~\cite{Tripathi2022}. 

The simplicity of our model allows a straightforward generalization to multi-component systems. However, the present model does not account for hydrodynamic interactions between the swimmers, which becomes significant in a dense suspension. Recent studies of tracer diffusion in a micro-swimmer suspension showed that the strong hydrodynamic forces induce correlations among the swimmers, leading to even faster diffusion of the tracer, which increases nonlinearly with the swimmer concentration~\cite{Morozov2022,Cheng2016}. Furthermore, anisotropic tracers exhibit more complex dynamics due to the coupling of  their translational and rotational motions~\cite{Morozov2022,cheng2016A}.

In summary, the present study provides simple, easy-to-calibrate relations between the motility of large tracer particles and the physical properties of a single nanoswimmer, specifically its dipole moment, and dynamical timescales. Knowledge of the physical characteristics of swimmers in these complex environments will be valuable in technological applications of nanomachines, for example, in facilitating drug delivery~\cite{Sanchez2021, Gao2014,Zhang2021}.

\bibliography{Ref}
 \newpage
 \clearpage

\setcounter{equation}{0}
\setcounter{figure}{0}
\setcounter{table}{0}
\setcounter{page}{1}
\makeatletter
\renewcommand{\theequation}{S\arabic{equation}}
\renewcommand{\thefigure}{S\arabic{figure}}
\renewcommand{\thetable}{S\arabic{table}}  
\setcounter{section}{0}
\renewcommand{\thesection}{S-\Roman{section}}

\section*{Supplementary Materials}
\noindent \textbf{Force on the tracer due to a single force dipole.}
Consider a spherical tracer whose center is at the origin, and a force dipole located at a distance $\rb$ from the tracer. The force dipole consists of two Stokeslets of equal but opposite strength $\fb=f\hfb$, separated by a displacement $\lbs =\lsS\heb$. The strength of the force dipole is described by dipole moment tensor $\mb =(f \lsS)\heb_j\hfb_k$. The velocity field generated by the dipole is
\begin{equation}
\label{eq:VelFDM}
 \vfd(\rb) = \frac{\hrb}{8\pi\mu}\left[-\frac{\fb\vdot\lbs}{r^2}+\frac{3(\fb\vdot\rb)(\lbs\vdot\rb)}{r^4}\right] +\frac{(\fb\times\lbs)\times\hrb}{8\pi\mu  r^2}~.
\end{equation}
 To calculate the force on the tracer, we apply Faxen's law
 \begin{equation}
 \label{eq:Faxen}
     \Fb = \left.\left[6\pi\mu \Rtr\left(1+\frac{a^2}{6}\nabla^2\right)\vfd(\rb)\right]\right\rvert_{\rb=0}~.
 \end{equation}
Substituting Eq.~(\ref{eq:VelFDM}) in (\ref{eq:Faxen}), we obtain the force
\begin{equation}
\label{eq:FSphM}
    \Fb = \Fb_{\Str} + \Fb_{\Rot}~,
\end{equation}
a sum of the contributions arising from the stresslet and rotlet, \begin{align}
\label{eq:FcolSS}
\Fb_{\Str} &= \frac{3\Rtr m}{4r^2}\Bigg[\hrb\l(-\heb\vdot\hfb +3(\heb\vdot\hrb)(\hfb\vdot\hrb)\r)+ \frac{\Rtr^2}{r^2}
\nonumber \\&\Big[\hrb\l(\heb\vdot\hfb  -5(\heb\vdot\hrb)(\hfb\vdot\hrb)\r) + \heb(\hfb\vdot\hrb)+ \hfb(\heb\vdot\hrb)\Big]\Bigg]~,
\nonumber\\
    \Fb_{\Rot} &= \frac{3\Rtr m}{4r^2}\left[ \hfb(\heb\vdot\hrb) - \heb(\hfb\vdot\hrb)\right]~.
\end{align}\\

\noindent \textbf{Brownian motion of a single force dipole}.~ Due to their nanometric size, force dipoles experience strong thermal forces exerted by the surrounding medium. This gives rise to the stochastic rotation of the dipole and force orientations $\heb$ and $\hfb$, which vary on a timescale defined by rotational diffusion. We study the rotational motion within Langevin's framework which captures the dynamics of the dipole and force orientations as~\cite{doi1988, doi2013}
\begin{align}
\label{eq:Rotdiff}
    \frac{d\hub}{dt} = \Omegab\times\hub~,
\end{align}
where $\Omegab$ is the thermally-induced rotational rate characterized by the moments,
\begin{align}
    \lang\Omegab(t)\rang = 0, ~~~ \lang\Omegab_\alpha(t)\Omegab_\beta(t')\rang = 2\Drs\delta_{\alpha\beta}\delta(t-t'),
\end{align}
and $\Drs$ is the rotational diffusion coefficient. This leads to a Fokker-Plank equation, similar to the spatial diffusion equation,
\begin{align}
    \frac{\partial P(\hub)}{\partial t} = \nabla^2_{\hub}P(\hub),
\end{align}
 with an additional constraint that, being a unit vector, $\hub$ rotates on the surface of a unit sphere, $|\hub|^2=1$. The above equation is solved using spherical coordinates with $\hub$ pointing in the radially outward direction ($\hub=\hrb$), and the solution is the linear superposition of spherical harmonics ~\cite{doi2013,berne2013}.
 
 A useful quantity is the conditional probability $G(\hub, t; \hubz,0)\dd{\hubz} \dd{\hub}$ that defines the probability of finding the orientation $\hub$---around the solid angle $\dd{\hub}$---which was initially at $\hubz$ within $\dd{\hubz}$. This Green function can be expressed as
 \begin{equation}
 \label{eq:JProb}
    G(\hub, t| \hubz,0) = K(\hub, t| \hubz,0)P(\hubz)~,
\end{equation}
 where $K(\hub, t| \hubz,0)$ is the transition probability, defined as 
 \begin{align}
 \label{eq:Probe}
     K(\hub, t| \hubz,0) = \sum_{l,m}\exp\l[-l(l+1)\Drs ~t\r]Y_{lm}(\hubz)Y^*_{lm}(\hub)~,
 \end{align}
 and $P(\hubz)$ is the probability distribution of initial orientations. For a dipole in thermal equilibrium, $P(\hubz) = 1/(4\pi)$. $Y_{lm}$ is a spherical harmonics of degree $l$ and order $m$, 
\begin{align}
\label{eq:Ylm}
    Y_{lm}(\theta, \phi) = (-1)^m\sqrt{\frac{2l+1}{4\pi}\frac{(l-m)!}{(l+m)!}}P_l^m(\cos\theta)e^{im\phi}~,
\end{align}
and $P_l^m(\cos\theta)$ is the associated Legendre function~\cite{arfken2005}.

Using Eq.~(\ref{eq:JProb}), we calculate the average of any arbitrary function of the orientations  $O(\hub, \hubz)$ as 
\begin{equation}
\label{Eq:AvgOu}
 \lang O(\hub, \hubz)\rang= \int \dd{\hubz} \int\dd{\hub} ~O(\hub, \hubz)G(\hub, t| \hubz,0)~.
\end{equation}
Since the moment of a function of $\hub$ corresponds to averaging at the same time, the transition probability reduces to unity, and the average $\lang O(\hub)\rang$ can be written as
 \begin{align}
 \label{eq:Moment}
   \lang O(\hub)\rang &= \int\dd{\hub} ~O(\hub)P(\hub) \nonumber \\
   &=  \frac{1}{4\pi}\int_0^\pi\dd{\theta} \sin\theta \int_0^{2\pi}\dd{\phi} ~O(\theta, \phi).
 \end{align}
 As an example, consider $O(\hub) = u_z^2 = \cos^2\theta$. Substituting in Eq.~(\ref{eq:Moment}), we find $\lang u_z^2\rang = 1/3$. 
 
 To calculate the autocorrelation $\lang O(\hub, \hubz)\rang$, notice that the spherical harmonics form an orthonormal basis set. Thus, we express $O(\hub, \hubz)$ as a linear superposition, $O(\hub, \hubz)= \sum_{l,m}\sum_{l'm'}c_{lm}d_{l'm'}Y^*_{lm}(\hub)Y^*_{l'm'}(\hubz)$. Using this expression and the orthonormality of $Y_{lm}$, one can easily calculate the autocorrelation. 
 For example, consider $O(\hub, \hubz)=\hub_z(t)\hubzz(0)=\cos\theta\cos\theta_0$. 
 Using spherical harmonics we write 
 $\hub_z(t)\hubzz(0) = \frac{4\pi}{3} Y^*_{10}(\hub) Y_{10}(\hubz)$. Now substituting this expression in Eq.~(\ref{Eq:AvgOu}) and using the orthonormality condition for the spherical harmonics, we obtain $\lang\hub_z(t)\hubzz(0)\rang = \third\exp(-2\Drs t)$. We employ the same procedure to evaluate the following moments and correlations required for further calculations,
\begin{align}
\label{eq:AvgU24}
&\lang \hub \rang = 0~; \qquad \lang \huba(t)\hubb(0)\rang = \frac{\delta_{\sal\sbe}}{3}e^{-2D_r t}~; \nonumber\\
&\langle\hub_\alpha(t) \hub_\beta(t)\hub_\mu(0)\hub_\nu(0)\rangle = \nonumber \\
&\qquad\qquad\begin{cases}
\frac{1}{9}\l[1+\frac{4}{5}e^{-6\Drs t}\r] &\text{if~ $\alpha=\beta=\mu=\nu$}\\
\frac{1}{9}\l[1-\frac{2}{5}e^{-6\Drs t}\r] &\text{if~ $\alpha=\beta$, $\mu=\nu$}\\
\frac{1}{15}e^{-6\Drs t}\ &\text{if ~$\alpha=\mu(\nu)$, $\beta=\nu(\mu)$}\\
 0 \ &\text{otherwise}.
\end{cases}
\end{align}

\noindent \textbf{Conformational motion of force dipoles}.~
For nanoscale swimmers such as active proteins and artificial nanomotors, the interaction forces are comparable to the thermal ones. Thus, the two Stokeslets comprising the dipole fluctuate, rendering the dipole moment, $m(t) = F(t)\lsS(t)$, a dynamic quantity. Here, we consider a model for dipole moment fluctuations induced by chemical reactions. Each chemical reaction excites a burst that persists for an average time $\tauc$. The dipole remains idle between two consecutive bursts which occur with an average period $\taup$. For simplicity, we consider bursts of constant magnitude $m$, allowing us to treat the dipole moment fluctuation as a two-state Markov process with the moment values $\{m, 0\}$. We calculate the statistics of dipole moment fluctuation by considering the evolution of transition probabilities $P(j,t|i,0)$ between these two states using the master equation~\cite{Bala2020},
\begin{equation}
   \frac{\d \Pb}{\d t}  = W\Pb~.
\end{equation}
Here, $\Pb(t)$ denotes a 2D column vector whose elements are the transition probabilities $P(j,t|i,0)$, and $W$ is the $2\times 2$ transition matrix
\begin{equation}
  W=  
\begin{pmatrix}
-\lambda_1 &\lambda_2\\
\lambda_1 & -\lambda_2
\end{pmatrix}~,
\end{equation}
where $\lambda_1 = 1/\tauc$ is the mean transition rate from the moment state $m(t) = m$ to $m(t)=0$, and $\lambda_2 = 1/(\taup-\tauc)$ is the mean transition rate for the reverse transition. The solution of the master equation is $\Pb(t) = e^{Wt}\Pb(0)$, where the elements of $e^{Wt}$ are the transition probabilities,
\begin{align}\label{eq:TransProb}
    &P(m, t|m) = \frac{\lmp+\lmc e^{-\lambda t}}{\lambda}~; ~~~  P(m, t|0) = \frac{\lmp(1- e^{-\lambda t})}{\lambda}~,\nonumber\\
    &P(0, t|m) = \frac{\lmc(1- e^{-\lambda t})}{\lambda}~; ~~~ P(0, t|0) = \frac{\lmc+\lmp e^{-\lambda t}}{\lambda}~,
\end{align}
with $\lambda= \lmc+\lmp$. The stationary probability can be obtained as the asymptotic limit ($t\to \infty$),
\begin{equation}\label{eq:StatProb}
    P(m) = \frac{\lmp}{\lambda}~; ~~~  P(0) = \frac{\lmc}{\lambda}~.
\end{equation}
The dipole-moment autocorrelation is
\begin{equation}
    \lang m(t) m(0)\rang = \sum_i\sum_j m_i m_j P(j, t|i)P(i)~,
\end{equation}
where summations are over the two states $\{m,0\}$ of the Markov process. Using Eqs.~(\ref{eq:TransProb}, \ref{eq:StatProb}), we obtain the moment autocorrelation,
\begin{equation}
    \lang m(t) m(0)\rang = \frac{m^2\lmp^2}{\lambda^2} + \frac{\lmc\lmp m^2}{\lambda^2}\,e^{-\lambda t}~.
\end{equation}
Now, substituting the expression for transition rates, $\lmc=1/\tauc$ and $\lmp=1/(\taup-\tauc)$, we obtain
\begin{equation}
\label{eq:MCorr}
    \lang m(t)m(0) \rang =
   m^2 b  \l[b + \l(1- b \r) e^{-t/\taum} \r]~,
\end{equation}
where $b = \tauc/\taup$ is the relative fraction of bursts and $\taum = \tauc(1-b)$ is the timescale of dipole moment fluctuations.\\

\noindent \textbf{Diffusion of macroscopic tracer in a suspension of small particles}.~
Consider a collection of force dipole situated at positions $\{\rb_i \}$ with random dipole and force orientations $\{\heb_i\}$ and $\{\hfb_i\}$. Ignoring the interactions among the dipoles, the linearity of stokes flow allows us to calculate the total force on a tracer as the sum of forces exerted by individual dipoles,
\begin{equation}
    \Fb^{\sT} = \Fb^{\sT}_{\Str} + \Fb^{\sT}_{\Rot}~,
\end{equation}
where $ \Fb^{\sT}_{\Str}$ and $ \Fb^{\sT}_{\Rot}$ are the forces arising due to the stresslet and rotlet parts of the swimmers velocity fields. These forces are calculated as 
\begin{align}
\Fb^{\sT}_{\Str} &=\sum_i \Fb^{(i)}_{\Str}=\nonumber\\ &\frac{3\Rtr}{4}\sum_i^N\frac{m_i\hrb_i}{r_i^2}\left[-(\heb_i\vdot\hfb_i) +3(\heb_i\vdot\hrb_i)(\hfb_i\vdot\hrb_i) \right]\nonumber\\
&+\frac{3\Rtr^3}{4}\sum_i^N\frac{m_i}{r_i^4}\Big[\hrb_i\l((\heb_i\vdot\hfb_i) -5(\heb_i\vdot\hrb_i)(\hfb_i\vdot\hrb_i)\r)\nonumber \\ &+ \heb_i(\hfb_i\vdot\hrb_i)+ \hfb_i(\heb_i\vdot\hrb_i)\Big]~,\\
\label{eq:FcolRRT}
\Fb^{\sT}_{\Rot} 
    &=\sum_i\Fb^{(i)}_{\Rot}= \frac{3\Rtr}{4}\sum_i^N\frac{m_i}{r_i^2}\l[\hfb_i(\heb_i\vdot\hrb_i)-\heb_i(\hfb_i\vdot\hrb_i)\r]~.
\end{align}

We now divide the force due to a single dipole into axisymmetric~($\heb=\hfb$) and transverse~($\heb\perp\hfb$) components, $\fb =\fpara\heb+\fperp \hfb$.  We first consider the axisymmetric case, where only the stresslet contributes to the force. The rotlet contribution vanishes, as can be seen from Eq.~(\ref{eq:FcolRRT}). Therefore the overall force on the tracer is
\begin{align}
\label{eq:Case1T}
\Fpara^{\sT} &= \frac{3\Rtr}{4}\sum_i^N\Bigg[\frac{\mparai\hrb_i}{r_i^2}\l(-1 +3(\heb_i\vdot\hrb_i)^2 \r) \nonumber \\
&+\frac{\Rtr^2\mparai}{r_i^4}\l(\hrb_i\l(1-5(\heb_i\vdot\hrb_i)^2\r) + 2\heb_i(\heb_i\vdot\hrb_i)\r)\Bigg]~,
\end{align}
where $\mpara = \fpara\lsS$, is the axisymmetric component of the dipole moment. To proceed further, we replace the summation over dipoles by integral using field-point notation and rewrite $\Fpara^{\sT}$ as
\begin{align}
  \Fpara^{\sT}(t)&=\frac{3\Rtr}{4}\int_{\Rtr} \dd{\rb} \sum_i^N\delta(\rb-\rb_i)\mparai(t) \Bigg[\frac{\hrbs}{r^2}\Big(-1 + \nonumber \\
  &3(\heb_i\vdot\hrbs)^2\Big) +
\frac{\Rtr^2}{r^4}\l(\hrbs\l(1-5(\heb_i\vdot\hrbs)^2\r)+ 2\heb_i(\heb_i\vdot\hrbs)\r)\Bigg],
\end{align}
where a lower cutoff to the integral has been introduced in the integration to account for the finite size of the tracer. The Force $\Fb^{\sT}$ is stochastic due to the conformational and rotation motion of the dipole as described above. Measuring the mean and autocorrelation of the force requires averaging over dipole positions and orientations and dipole moments. Assuming that the orientations evolve independently of the positions, we find the total force
\begin{align}
\lang\Fpara^{\sT}(t)\rang&=\frac{3\Rtr}{4}\int_{\Rtr} \dd{\rb} C(\rb)\lang \mpara(t)\rang \Bigg[\frac{\hrbs}{r^2}\lang-1 +3(\heb\vdot\hrbs)^2\rang \nonumber \\
&+\frac{\Rtr^2}{r^4}\lang\hrbs\l(1-5(\heb\vdot\hrbs)^2\r)+ 2\heb(\heb\vdot\hrbs)\rang\Bigg]~,
\end{align}
where $C(\rb) = \lang\sum_i^N\delta(\rb-\rb_i)\rang$ is the average concentration of force dipoles. 
Now, consider the first term in the average, $\lang -1 +3(\heb\vdot\hrbs)^2\rang=-1+3\sum_{\alpha,\beta}\lang\heb_{\alpha}\heb_{\beta}\rang\hrb_{\alpha}\hrb_{\beta}$. Using Eq.~(\ref{eq:AvgU24}), we obtain $\lang -1 +3(\heb\vdot\hrbs)^2\rang=-1+\sum_{\alpha,\beta}\delta_{\alpha_\beta}\hrb_{\alpha}\hrb_{\beta} = -1+\sum_{\alpha}\hrb_\alpha^2 = 0$. 
Similarly one can show that $\lang\l(1-5(\heb\vdot\hrbs)^2\r)+ 2\heb(\heb\vdot\hrbs)\rang =0$. Thus, the mean force generated by the suspension vanishes, as one expects from symmetry.

Next, we calculate the autocorrelation,
\begin{align}
  \lang\Fpara^{\sT}(t)\Fpara^{\sT}(0)\rang =  \sum_i\sum_j\lang\Fb^{(i)}_{\Str}(t)\Fb^{(j)}_{\Str}(0)\rang
\end{align}
Since the hydrodynamic coupling between the dipoles is neglected, the average over orientations vanishes for $i\neq j$. Again writing in field-point notation,
\begin{align}
\label{eq:FtotSSS}
 \big<\Fpara^{\sT}(t) &\Fpara^{\sT}(0)\big>=\frac{9\Rtr^2}{16}\int_{\Rtr} \dd{\rb} C(\rb)\lang \mpara(t)\mpara(0)\rang  \Bigg[\nonumber \\
 &\frac{\hrb\hrb}{r^4}\lang\l(-1 +3(\heb(t)\vdot\hrbs)^2\r)\l(-1 +3(\heb(0)\vdot\hrbs)^2\r)\rang \nonumber\\
 &+\frac{\Rtr^4}{r^8}\Big<\Big(\hrbs\l(1-5(\heb(t)\vdot\hrbs)^2\r)+ 2\heb(t)(\heb(t)\vdot\hrbs)\Big)\nonumber\\
 &\qquad\quad\Big(\hrbs\l(1-5(\heb(0)\vdot\hrbs)^2\r)+ 2\heb(0)(\heb(0)\vdot\hrbs)\Big)\Big> \nonumber \\
 &+\frac{2\Rtr^2\hrb}{r^6}\Big<\Big(\hrbs\l(1-5(\heb(t)\vdot\hrbs)^2\r)+ 2\heb(t)(\heb(t)\vdot\hrbs)\Big)\nonumber\\
 &\qquad\qquad\Big(-1 +3\l(\heb(0)\vdot\hrbs\r)^2\Big)\Big>
 \Bigg].
\end{align}
To perform the orientational averages, notice that the Eq.~(\ref{eq:FtotSSS}) contains second and fourth-order correlations of the form $\langle\heb_\alpha(t)\heb_\beta(t)\rangle$ and $\langle\heb_\alpha(t)\heb_\beta(t)\heb_\mu(0)\heb_\nu(0)\rangle$. Thus, using Eq.~(\ref{eq:AvgU24}), we obtain
\begin{align}
  \lang\Fparaa^{\sT}(t) \Fparab^{\sT}(0)\rang = \frac{9\Rtr^2}{20}\lang \mpara(t)\mpara(0)\rang \,e^{-3t/\taur}\int_{\Rtr} \dd{\rb} C(\rb)\nonumber\\
  \Bigg[\frac{\hrb_\alpha\hrb_\beta}{r^4}-\frac{2a^2\hrb_\alpha\hrb_\beta}{r^6}+\frac{a^4}{3r^8}\l(\delta_{\alpha\beta}+\hrb_\alpha\hrb_\beta\r)\Bigg]~,
\end{align}
Where Greek indices denote the Cartesian components. Considering the homogeneous solution~($C(r)=c_0$) where dipoles are isotropically distributed throughout the solution, the integral over position only survives when $\alpha=\beta$. Thus,
\begin{align}
   \lang\Fparaa^{\sT}(t) \Fparab^{\sT}(0)\rang &= \delta_{\alpha\beta}\frac{9\Rtr^2}{20}c_0\lang \mpara(t)\mpara(0)\rang \,e^{-3t/\taur} \nonumber\\
   &\int_{\Rtr} \dd{\rb}\Bigg[\frac{\hrb_\alpha^2}{r^4}-\frac{2a^2\hrb_\alpha^2}{r^6}+\frac{a^4}{3r^8}\l(1+\hrb_\alpha^2\r)\Bigg].
\end{align}
We use spherical coordinate system to calculate the above integral. For a finite suspension box of size $L$, The upper limit of the integration restricted to $r=L/2$. Thus we obtain 
\begin{align} 
\label{eq:FCorrL}
\lang\Fparaa^{\sT}(t) \Fparab^{\sT}(0)\rang = \delta_{\alpha\beta}&\frac{\pi \Rtr}{5}c_0 \lang \mpara(t)\mpara(0)\rang e^{-\frac{3t}{\taur}}\nonumber \\ &  \Bigg[2-\frac{3\Rtr}{L/2} +\frac{2\Rtr^3}{(L/2)^3} -\frac{\Rtr^5}{(L/2)^5}\Bigg]~.
\end{align}
For an infinite system, this simplifies to
\begin{equation}
\label{eq:FFCorrA}
\lang\Fparaa^{\sT}(t) \Fparab^{\sT}(0)\rang =  \delta_{\alpha\beta}\frac{2\pi \Rtr}{5}c_0 \lang \mpara(t)\mpara(0)\rang\exp\left(-\frac{3t}{\taur}\right)~.
\end{equation}\\

Similarly, We consider the case of transverse dipole~($\heb\vdot\hfb = 0$). In this case, both the stresslets and the rotlets contribute to the total force,
\begin{align}
\label{eq:integral2}
\Fperp^{\sT}(t) &= \frac{3\Rtr}{4}\int\dd{\rb}\sum_i^N\delta(\rb-\rb_i) \mperpi(t) \Bigg[\frac{3\hrbs}{r^2} (\heb_i\vdot\hrbs)(\hfb_i\vdot\hrbs)\nonumber \\
&+\frac{\Rtr^2}{r^4}\l(-5\hrbs(\heb_i\vdot\hrbs)(\hfb_i\vdot\hrbs)+ \heb_i(\hfb_i\vdot\hrbs)+ \hfb_i(\heb_i\vdot\hrbs)\r)\nonumber \\ &+\frac{1}{r^2}\l( \hfb_i(\heb_i\vdot\hrbs)-\heb_i(\hfb_i\vdot\hrbs)\r)\Bigg].
\end{align}.

By performing the average over orientations and moments as in the case of the axisymmetric dipole, we obtain the force autocorrelation
\begin{equation}
\label{eq:FFCorrT}
 \lang\Fperpa^{\sT}(t) \Fperpb^{\sT}(0)\rang =  \delta_{\alpha\beta}\frac{4\pi}{5} c_0 \Rtr\lang \mperp(t)\mperp(0)\rang\exp\left(-\frac{3t}{\taur}\right).
\end{equation}
Combining Eqs.~(\ref{eq:FFCorrA}, \ref{eq:FFCorrT}) and using Eq.~(\ref{eq:MCorr}), we obtain the expression for the autocorrelation for the total force as reported in the manuscript.\\

\noindent \textbf{Brownian dynamic simulation of tracer motion in nano-swimmer suspensions}.~ 
We consider a suspension of nano-dipoles randomly distributed in a cubic box, with positions and orientations drawn from a uniform random distribution. In the following, we adopted physical units where the thermal energy, dipole size, and rotation diffusion time are unity. In the above units, the viscosity of the suspension is $\eta\approx 2/\pi$, and the box size is $L=150$.

We model the translational motion of dipole as
\begin{align}
\label{eq:dipoleM}
\frac{d\rb_i}{d t} = \sqrt{2\Dfd}~\nth(t)
\end{align}
where $\Dfd$ is the thermal diffusion coefficient of the dipole, and $\nth$ is uncorrelated Gaussian noise of zero mean and unit variance. The rotational motion is simulated using Eq.~(\ref{eq:Rotdiff}). 
The effect of moment fluctuations is modeled as a two-state Markov process described in the main text. The parameters related to the moment fluctuation is $m=10$, $\tauc=0.1$ and $\taup=0.2$. To generate a numerical realization of the Markov process, we assume that the dipole has the initial moment $m(0)=m$. We then calculate the probability $P(m, t|m)$ (Eq.~(\ref{eq:TransProb})) and compare it with an uniform random number $RN_1$ in [0,1]. If $P(m, t|m) > RN_1 $, the dipole remains in state $m(t)=m$, else it jump to the state $m(t)=0$. On the other hand, if dipole is in state $m(t)=0$, we calculate the probability $P(m, t|0)$ and generate another uniform random number $RN_2$ in [0,1]. Now, if $P(m, t|0) > RN_2$ dipole jump to state $m(t)=m$, else it remain in state $m(t)=0$. Iterating this procedure generates a time series for the 2-step Morkov process used in our simulation.

We consider a tracer particle of radius $a=50$, initially situated at the center of the box. The hydrodynamic interaction between the dipoles and the tracer is evaluated using Eq.~(\ref{eq:Faxen}), and the dynamics of the tracer's center are simulated using the following equation
\begin{align}
\label{eq:TracerM}
    \frac{d\Rb}{d t} = \sqrt{2\Dth}~\nth(t) + \Fb^{\sT}(t)/\gamma.
\end{align} 
We solve the Eqs.~(\ref{eq:Rotdiff}, \ref{eq:dipoleM}, \ref{eq:TracerM}) using the Euler scheme
\begin{align}
 \heb(t+dt) &= \heb(t) + \sqrt{dt}~ \Omegab_{\mathrm{G}}(t)\times \heb(t)~,\nonumber \\
 \rb_i(t+dt) &= \rb_i(t) + \sqrt{2~dt~\Dfd}~\mathbf{Y}_{\mathrm{fd}}~, \nonumber \\
 \Rb(t+dt) &= \Rb(t) + \sqrt{2~dt~\Dth }~\mathbf{Y}_{\mathrm{T}} + dt~\Fb^{\sT}(t)/\gamma~,
\end{align}
where $\Omegab_{\mathrm{G}}$ $\mathbf{Y}_{\mathrm{fd}}$, and $\mathbf{Y}_{\mathrm{T}}$ are uncorrelated Gaussian random variables of unit mean and zero variance, and $dt$ is the time step of the simulation. The thermal diffusion coefficients for the tracer and dipoles are $\Dth=\kBT/(6\pi\eta \Rtr)\approx0.002$ and $\Dfd=\kBT/(3\pi\eta\lsS) \approx 0.2$. Thus, by measuring the tracer position and the force on the tracer, we calculate the power spectrum of force and diffusion coefficient and compare them to the analytical results, as discussed in the manuscript.

\end{document}